\documentclass[A4,12pt]{article}
\usepackage[T1]{fontenc}
\usepackage[utf8]{inputenc}
\usepackage[brazil]{babel}
\usepackage{lineno,hyperref}
\usepackage{subfigure}
\usepackage{graphicx}
\usepackage{mathrsfs}
\usepackage{algorithm}
\usepackage{algpseudocode}
\algtext*{EndProcedure}    
\begin{document}

\title{Two-dimensional Mesh Generator in Generalized Coordinates Implemented in Python}


\author{
Gustavo Taiji Naozuka, Saulo Martiello Mastelini \\
	Departamento de Computação \\ 
	Universidade Estadual de Londrina\\
	gtnaozuka@gmail.com,saulomastelini@gmail.com \\ \\
Eliandro Rodrigues Cirilo, Neyva Maria Lopes Romeiro, Paulo Laerte Natti \\
	Departamento de Matemática \\
	Universidade Estadual de Londrina\\
	nromeiro@uel.br,ercirilo@uel.br,plnatti@uel.br}
\maketitle


\noindent \textbf{Abstract} \\
Through mathematical models, it is possible to turn a problem of the physical domain to the computational domain. In this context, the paper presents a two-dimensional mesh generator in generalized coordinates, which uses the Parametric Linear Spline method and partial differential equations. The generator is automated and able to treat real complex domains. The code was implemented in Python, applying the Numpy and Matplotlib libraries to matrix manipulations and graphical plots, respectively. Applications are made for monoblock meshes (two-dimensional shape of a bottle) and multiblock meshes (geometry of Igapó I lake, Londrina, Paraná, Brazil).

\noindent \textbf{keyword}:Automated two-dimensional mesh generator, Parametric Linear Spline, generalized coordinates, Python language, applications.   \\ \\




\section{Introduction}

The modeling and simulation of natural phenomena using differential equations is an important tool for science. However, to represent physical structures, that is, the domain to be studied in a computational environment, the use of simpler data organization structures, such as matrices and vectors, tend not to be able to realistically represent the object of study. Thus, different techniques have been used for the representation of domains, such as structured and unstructured meshes, generalized or hybrid meshes, among other techniques, that have advantages and disadvantages in terms of flexibility and ability to represent objects under study (THOMPSON et al., 1977; THOMPSON et al., 1985; THOMPSONet al., 1998; MALISKA, 2004; CIRILO et al. 2006; KOOMULLIL et al, 2008;  LAIPING et al., 2013; SAITA et al., 2017; BELINELLI et al, 2020).

In this work, we looked for a simplified way to build a structured mesh for a complex domain, which was possible through a change in the coordinate system. This process allowed to convert a complex domain into a set of easily manipulated data (MALISKA, 2004). Mathematically, any geometry, described in Cartesian system, can be transformed into a generalized system, allowing better adaptation in computational modeling (CIRILO et al. 2006; MALISKA, 2004; ROMEIRO et al., 2011; PARDO et al., 2012; SAITA et al., 2017; ROMEIRO et al., 2017; CIRILO et al., 2018).

In this context, this work describes the creation and implementation of a two-dimensional mesh generator using transformation metrics in generalized coordinates. This procedure generates a domain mapped for mathematical manipulations, which describes any physical object, from a finite set of points. The mesh generator is encoded in \textit{Python} and uses the libraries \textit{numpy} and \textit{matplotlib} for manipulations and operations on matrices and graphical plots, respectively (CAI et al., 2005; SCIPY-NUMPY, 2020; SCIPY-MATPLOTLIB,2020).

The work is structured as shown below. Initially, transformation metrics are described in generalized coordinate theory. Then the mesh generator is developed. Finally, some examples of meshes obtained by the generator are presented.

\section{Generalized Coordinate}

For a computational methodology to be applied to a physical problem, it is necessary to discretize the domain of the problem, that is, to build a computational mesh that can represent the studied geometry and, thus, obtain the values of interest.

The discretization of the physical domain can be carried out according to a structured or unstructured mesh. Unstructured meshes are more adaptable than structured meshes, especially in problems with complex geometries (MALISKA, 2004; FORTUNA, 2012). However, the major disadvantage of unstructured meshes is the difficulty of ordering the elements, which implies a variation in the size of the diagonals of the coefficient matrix and the additional cost of using memory, making it difficult to apply numerical methods to obtain the solution of linear systems. Therefore, the domain discretization in this work will be performed using structured meshes.

As for the coordinate system, in general, the problem domain is discretized according to the Cartesian coordinate system because it is simpler. However, for problems with complex geometry, it is convenient to adopt another coordinate system, due to the fact that the Cartesian coordinate system leads to a poor adaptation of the border, since the physical domain does not always coincide with the domain of the Cartesian mesh. To solve the problem, the generalized coordinate system will be used. In generalized coordinates, the computational mesh coincides with the geometry of the problem and computational treatment becomes more appropriate. Other reasons that justify the use of generalized coordinates in the discretization of computational meshes are the simplicity in programming computational codes to solve complex problems and the ease in developing generic methodologies.

In the following, some important concepts about generalized coordinate theory will be presented

\subsection{Transformation Metrics}

The Cartesian system $(x, y)$ is called the physical domain and the generalized system $(\xi, \eta)$ is called the transformed domain or computational domain. 

The transformation from a non-trivial geometry described in a Cartesian coordinate system to a generalized coordinate system involves transformation metrics, or mathematical relations, that can accurately describe the transformed data.

The mapping of irregular or regular geometries written in Cartesian coordinates $(x, y)$, Fig. 1a, is performed numerically for regular geometries written in generalized coordinate system $(\xi, \eta)$, Fig. 1b.

\begin{figure}[!ht]
\begin{center}
\caption{Coordinate systems}
\subfigure[Cartesian coordinate]{\includegraphics[height=4.4cm,width=5.4cm]{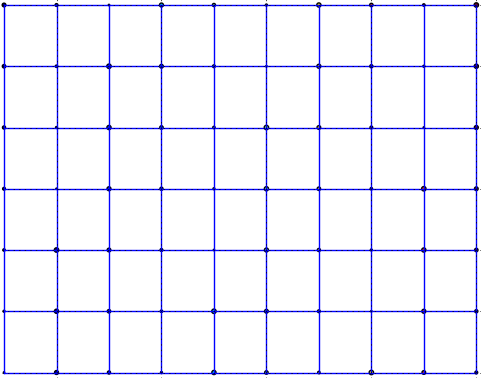}}
   \label{fig2a}\qquad \qquad
\subfigure[Generalized coordinate]{\includegraphics[height=4.4cm,width=5.4cm]{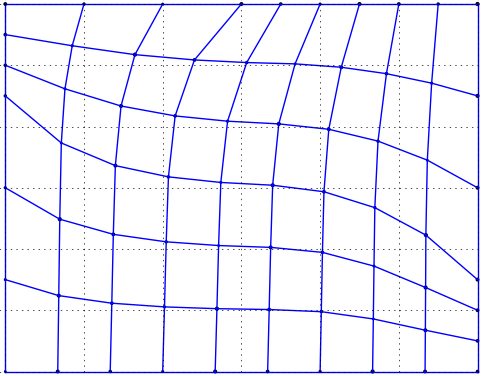}}
    \label{fig2b}
\end{center}
\center{{\bf{Source:}} The Authors}
\label{fig2}
\end{figure}

Since the transformed domain is regular, for convenience, unitary normalization of elementary volumes is assumed, that is, $\Delta\xi=\Delta\eta=1$. In this way, even if in the physical plane the coordinated lines assume arbitrary spacing, in the computational plane the dimensions are fixed to the unit.

To obtain the $(\xi, \eta)$ system, the following generating equations are used
\begin{eqnarray}
	\label{eq1}
		\xi = \xi(x,y), \\
	\label{eq1a}
		\eta = \eta(x,y).
\end{eqnarray}

The transformation metrics, based on differentials of the Eqs. (\ref{eq1}) and (\ref{eq1a}), are
\begin{eqnarray}
	\label{eq2}
		&d\xi = \xi_{x}dx + \xi_{y}dy, \\
	\label{eq2a}
		&d\eta = \eta_{x}dx + \eta_{y}dy,
\end{eqnarray}

\noindent 
or, in the matrix form
\begin{equation}
	\left(
		\begin{array}{c}
			d \xi\\
			d \eta
		\end{array}
	\right)
	=
	\left(
		\begin{array}{cc}
			\xi_{x} 	& \xi_{y}\\
			\eta_{x}	 & \eta_{y}
		\end{array}
	\right)
	\left(
		\begin{array}{c}
			d x\\
			d y
		\end{array}
	\right),
	\label{eq3}
\end{equation}

\noindent
which can also be written as
\begin{equation}
	d^{t} = Ad^{f},
	\label{eq4}
\end{equation}

\noindent
where $\xi_{x}$, $\eta_{x}$, $\xi_{y}$, and $\eta_{y}$ denoting partial derivatives, $d^{t}$ and $d^{f}$ represent, respectively, the differentials in the transformed and physical domain, while $A$ is the transformation matrix between the domains.

Starting from the assumption that it is possible to find a representation in a Cartesian coordinate system for a model described in generalized coordinates, then we admit the existence of the inverse of the equations Eqs. (\ref{eq1}) and (\ref{eq1a}), so
\begin{eqnarray}
	\label{eq9}
		&x = x(\xi,\eta), \\
	\label{eq9a}
		&y = y(\xi,\eta),
\end{eqnarray}

\noindent
and from differentials of Eqs. (\ref{eq9}) and (\ref{eq9a}) we have
\begin{equation}
	\left(
		\begin{array}{c}
			d x\\
			d y
		\end{array}
	\right)
	=
	\left(
		\begin{array}{cc}
			 x_{\xi} 	& x_{\eta}\\
			 y_{\xi} 	& y_{\eta}
		\end{array}
	\right)
	\left(
		\begin{array}{c}
			d \xi\\
			d \eta
		\end{array}
	\right),
	\label{eq5}
\end{equation}

\noindent
which can also be written as

\begin{equation}
	d^{f} = Bd^{t},
	\label{eq6}
\end{equation}

\noindent
where $x_{\xi}$, $x_{\eta}$, $y_{\xi}$, and $y_{\eta}$ denoting partial derivatives and $B$ is the transformation matrix between the physical and transformed domains.

Replacing Eq. (\ref{eq4}) in Eq. (\ref{eq6}), we get  $d^{f} = BAd^{f}$, so that $BA = I$, or equivalent $A = B^{- 1}$. Thus, the $A$ matrix becomes
\begin{equation}
	\left(
		\begin{array}{cc}
			\xi_{x} 	& \xi_{y}\\
			\eta_{x}	 & \eta_{y}
		\end{array}
	\right)
	=
	\left(
		\begin{array}{cc}
			\frac{y_{\eta}}{x_{\xi}y_{\eta} - x_{\eta}y_{\xi}} & \frac{-x_{\eta}}{x_{\xi}y_{\eta} - x_{\eta}y_{\xi}}\\
			\frac{-y_{\xi}}{x_{\xi}y_{\eta} - x_{\eta}y_{\xi}} & \frac{x_{\xi}}{x_{\xi}y_{\eta} - x_{\eta}y_{\xi}}
		\end{array}
	\right),
	\label{eq7}
\end{equation}

\noindent
where $J = (x_{\xi}y_{\eta} - x_{\eta}y_{\xi})^{-1}$ is called Jacobian of transformation (MALISKA, 2004).

\subsection{Two-dimensional Mesh Generation}

In this work, we opted for the use of elliptical partial differential equations (EPDE) as a method of generating two-dimensional meshes, since their solutions do not generate null Jacobian, and the lines $\xi$ or $\eta$ never intersect (THOMPSON, 1985; MALISKA, 2004). Thus, the governing equations for the generation of two-dimensional meshes in a domain, for example the domain illustrated in Fig. 2a, are
\begin{eqnarray}
	\label{eq13}
		&\nabla^{2}\xi = P(\xi,\eta), \\
	\label{eq13a}
		&\nabla^{2}\eta=Q(\xi,\eta),
\end{eqnarray}

\noindent
whose boundary conditions of the type \textit{Dirichlet} are expressed by 
$\xi = \xi_ {1}$ in $\Gamma_ {1}$ (left border), 
$\xi = \xi_{N}$ in $\Gamma_{3}$ (right border); 
$\eta = \eta_{1}$ in $\Gamma_{4}$ (lower border) and 
$\eta = \eta_{M}$ in $\Gamma_{2}$ (upper border). 
In particular, in Fig. 2a were used 
$N = 10$ and $M = 5$ , where $N$ is the number of lines from 
$\xi$ and $M$ is the number of lines from $\eta$.

\begin{figure}[!ht]
\begin{center}
\caption{Example of mesh in generalized coordinates}
\subfigure[Predefined points]{\includegraphics[height=4.4cm,width=5.4cm]{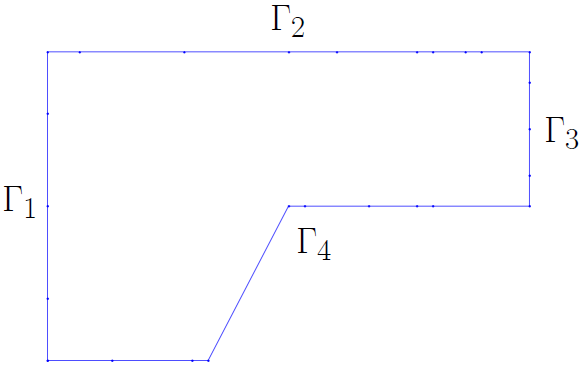}}
   \label{fig4a}\qquad \qquad
\subfigure[Mesh in generalized coordinates]{\includegraphics[height=4.4cm,width=5.4cm]{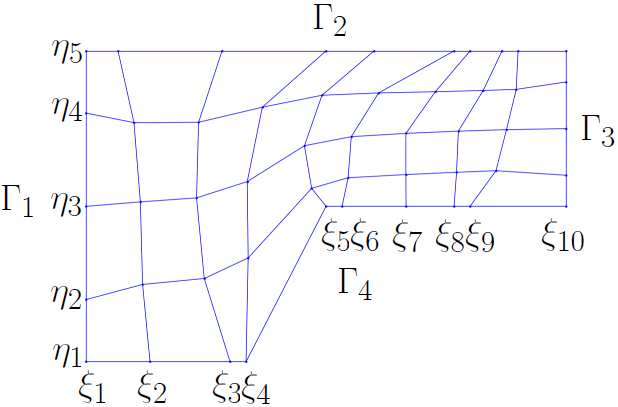}}
    \label{fig4b}
\end{center}
\center{{\bf{Source:}} The Authors}
\label{fig4}
\end{figure}

Solving the equations (\ref{eq13}) and (\ref{eq13a}), in relation to the $(x, y)$ coordinates, using the transformation metrics of the $(\xi, \eta)$ system, then the coordinated lines can be generated in the directions $\xi$ and $\eta$, inside the computational mesh, Fig. 2b, as in any other geometry, through the equations
\begin{eqnarray}
	\label{eq25}
	\alpha x_{\xi\xi} + \gamma x_{\eta\eta} - 2\beta x_{\xi\eta} + \frac{1}{J^{2}}(Px_{\xi} + Qx_{\eta}) = 0, \\
	\label{eq26}
	\alpha y_{\xi\xi} + \gamma y_{\eta\eta} - 2\beta y_{\xi\eta} + \frac{1}{J^{2}}(Py_{\xi} + Qy_{\eta}) = 0,
\end{eqnarray}

\noindent
where $x$ and $y$ are the Cartesian coordinates of the physical domain, $\xi$ and $\eta$ are the generalized coordinates of the computational domain, $P$ and $Q$ are the source functions, $J$ is the Jacobian of transformation, and
\begin{eqnarray}
	\label{eq26a}
		\alpha = x^{2}_{\eta} + y^{2}_{\eta} , \\
	\label{eq25b}
		\beta = x_{\xi}x_{\eta} + y_{\xi}y_{\eta}, \\
	\label{eq26b}
		\gamma = x^{2}_{\xi} + y^{2}_{\xi}.
\end{eqnarray}

The numerical solution of elliptical EDPs (\ref{eq25}) 
and (\ref{eq26}), subject to initial and boundary conditions, provide the lines $\xi$ and $\eta$, which generate the computational mesh (DE BORTOLI, 2000, MALISKA, 2004; CIRILO et al., 2006; FORTUNA, 2012). For convenience, Eqs. (\ref{eq25}) and (\ref{eq26}) can be written using a generic $\phi$ variable as
\begin{equation}
	\alpha\phi_{\xi\xi} + \gamma\phi_{\eta\eta} - \alpha\beta\phi_{\xi\eta} + \frac{1}{J^{2}}(P\phi_{\xi} + Q\phi_{\eta}) = 0.
	\label{eq27}
\end{equation}

\begin{figure}[!ht]
\begin{center}
\caption{Index labeling}
  \includegraphics[height=4.4cm,width=5.4cm]{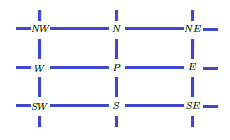}
	\label{fig5}
\end{center}
\center{{\bf{Source:}} The Authors}
\label{fig4}
\end{figure}

To approximate the derivatives in Eq. (\ref{eq27}), the finite difference method is used. The mesh nodes are labeled by the cardinal points $P$ (center), $E$ (east), $W$ (west), $N$ (north), $S$ (south), $NW$ (northwest ), $SW$ (southwest), $SE$ (southeast) and $NE$ (northeast), as in Fig. \ref{fig5}.

Approximating the derivative terms of Eq. (\ Ref {eq27}) by means of central differences, around $ P $, we obtain
\begin{eqnarray}
		\alpha(\frac{\phi_{E} - 2\phi_{P} + \phi_{W}}{\Delta\xi^{2}}) + \gamma(\frac{\phi_{N} - 2\phi_{P} + \phi_{S}}{\Delta\eta^{2}}) - 2\beta(\frac{\phi_{NE} - \phi_{NW} + \phi_{SW} - \phi_{SE}}{4\Delta\xi\Delta\eta}) + \nonumber\\ 
		\frac{1}{J^{2}}(P\frac{\phi_{E} - \phi_{W}}{2\Delta\xi} + Q\frac{\phi_{N} - \phi_{S}}{2\Delta\eta}) = 0,
\label{eq28}
\end{eqnarray}

\noindent
and regrouping the terms
\begin{eqnarray}
		-(2\alpha - 2\gamma)\phi_{P} + (\alpha + \frac{P}{2J^{2}})\phi_{E} + (\alpha - \frac{P}{2J^{2}})\phi_{W} + (\gamma + \frac{Q}{2J^{2}})\phi_{N} + \nonumber\\
		(\gamma - \frac{Q}{2J^{2}})\phi_{S} - \frac{\beta}{2}\phi_{NE} + \frac{\beta}{2}\phi_{NW} - \frac{\beta}{2}\phi_{SW} + \frac{\beta}{2}\phi_{SE} = 0.
\label{eq29}
\end{eqnarray}

Thus, the numerical solution of the two-dimensional mesh generation equations, Eqs. (\ref{eq25}) and (\ref{eq26}), written according to the generic $\phi$ variable, 
Eq. (\ref{eq29}), is given by
\begin{eqnarray}
		\phi_{P} = \frac{1}{A_{P}}(A_{E}\phi_{E} + A_{W}\phi_{W} + A_{N}\phi_{N} + A_{S}\phi_{S} + A_{NE}\phi_{NE} + A_{SE}\phi_{SE} + \nonumber\\
	\label{eq30}
		A_{NW}\phi_{NW} + A_{SW}\phi_{SW}),
\end{eqnarray}

\noindent
where
\begin{eqnarray}
	\begin{array}{lclcl}
		A_{P} = 2\alpha + 2\gamma, &\quad & A_{N} = \gamma + \frac{Q}{2J^{2}}, &\quad & A_{SE} = \frac{\beta}{2}, \\ \\
		A_{E} = \alpha + \frac{P}{2J^{2}}, &\quad & A_{S} = \gamma - \frac{Q}{2J^{2}}, &\quad & A_{NW} = \frac{\beta}{2}, \\ \\
		A_{W} = \alpha - \frac{P}{2J^{2}}, &\quad & A_{NE} = -\frac{\beta}{2}, &\quad & A_{SW} = -\frac{\beta}{2},
	\end{array}
\end{eqnarray}

\noindent
with derivatives present in $J$ approximated by central differences.

Next, the mesh generator in generalized coordinates is presented.

\section{Mesh Generator in Generalized Coordinate}

For the creation of an automated procedure for mapping a geometry in Cartesian coordinates, through a generalized coordinate system, we chose to use the programming language 
Python. 

The reason for this choice is due to the fact that Python is a general, free, open, and multiplatform language, which can allow the creation of extensions of the developed application, such as, for example, a graphical interface module, among others. 

Regarding performance, it is necessary to use extra libraries, specific to numerical computing applications, to achieve satisfactory results at run time. Because it is a scripting language, that is, interpreted, the performance of an application written in Python tends to be inferior when compared to a compiled language, as is the case of the Fortran language. In Python, the tools available for matrix manipulation and linear algebra are limited and not optimized for the type of application that was intended to be developed. However, because the adopted language is modular, it is possible to use routines and libraries in compiled languages, in order to optimize more complex operations, such as those used for domain transformations (CAI et al., 2005).

For the creation of the mesh generator, the following libraries were used:
\begin{itemize}
	\item Numpy - library dedicated to matrix and algebraic operations in general (SCIPY-NUMPY, 2020);
  \item Matplotlib - library used to generate graphs and manipulate data in graphical form (SCIPY-MATPLOTLIB, 2020).
\end{itemize}

The additional modules employed are also free for use.

\subsection{Description of the Algorithm Developed}

For the definition and creation of a geometry mapped in generalized coordinates, it starts from a initial set of points that describe the border, that is, the physical contour. In this work, the Parametric Linear Spline method was chosen to interpolate the set of border points under study. Parametric Linear Spline equations (BURDEN et al., 2015) are
\begin{equation}
	\label{eq33a}
		s^{x}_{i}(t) = x_{i-1}\frac{t_{i} - t}{t_{i} - t_{i-1}} + x_{i}\frac{t - t_{i-1}}{t_{i} - t_{i-1}}, 
\end{equation}
\begin{equation}
	\label{eq33b}
		s^{y}_{i}(t) = y_{i-1}\frac{t_{i} - t}{t_{i} - t_{i-1}} + y_{i}\frac{t - t_{i-1}}{t_{i} - t_{i-1}}, 
\end{equation}
$\forall t \in [t_{i-1},t_{i}]$, where $s$ interpolating spline curve and $t$ is the interpolated variable.

Having obtained the interpolating lines of the border, 
Eqs. (\ref{eq33a}) and (\ref{eq33b}), the number of partitions desired for each border must be defined. 

Then, points of the Splines curves are selected, which define the new border of the approximate geometry. To do this, an approach is used to position the points by weighted average of the components of the border, ie
\begin{equation}
	\label{eq34a}
		x_{i j} = p^{\Gamma_{1} \Gamma_{3}}_{x} (x_{0 j} + i \Delta^{x \xi}_{j-1}) + p^{\Gamma_{2} \Gamma_{4}}_{x} (x_{i 0} + j \Delta^{x \eta}_{i-1}), 
\end{equation}
\begin{equation}
	\label{eq34b}
		y_{i j} = p^{\Gamma_{1} \Gamma_{3}}_{y} (y_{0 j} + i \Delta^{y \xi}_{j-1}) + p^{\Gamma_{2} \Gamma_{4}}_{y} (y_{i 0} + j \Delta^{y \eta}_{i-1}), 
\end{equation} 

\vspace{0.3cm}

\noindent
where $\Delta^{x \xi}_{j} = \frac{x_{\xi j} - x_{0 j}}{\xi}$, $\Delta^{x \eta}_{i} = \frac{x_{i \eta} - x_{i 0}}{\eta}$, $\Delta^{y \xi}_{j} = \frac{y_{\xi j} - y_{0 j}}{\xi}$, $\Delta^{y \eta}_{i} = \frac{y_{i \eta} - y_{i 0}}{\eta}$, $i = 1,...,\xi$ e $j = 1,...,\eta$.

\vspace{0.3cm}
\noindent
The values of $p^{\Gamma_{1} \Gamma_{3}}_{x}$, $p^{\Gamma_{2} \Gamma_{4}}_{x}$, $p^{\Gamma_{1} \Gamma_{3}}_{y}$, and $p^{\Gamma_{2} \Gamma_{4}}_{y}$ indicate the weights on the left/right borders in $x$, top/bottom in $x$, left/right in $y$ and top/bottom in $y$ , respectively. The percentage weights in Eqs. (\ref{eq34a}) and (\ref{eq34b}) can better adjust the distribution of points in the domain, avoiding concentrations of points. In this way, the arrangement of the internal points is influenced by the edges. 

Finally, on the set of ordered pairs, the resolution of the equation (\ref{eq30}) is applied.
\vspace {0.3cm}

In summary, the implemented algorithm can be illustrated in this way:

\noindent
Step 1. Reading the input data: Read the number of partitions from the physical and transformed planes, the weights for calculating the weighted average and the border points.

\noindent
Step 2. Border interpolation: Define the border using the Parametric Linear Spline method and the border points previously read.

\noindent
Step 3. Calculate the points of the Spline lines, using the weighted average formulas, and define the new border points.

\noindent
Step 4. Application of transformation metrics: On the generated data, apply the transformation metrics to obtain the interior mesh curves.

\vspace {0.3cm}
The next two sections are intended to exemplify and test the algorithm. In the first applications, monoblock meshes are built (3 applications). Following a multiblock mesh is obtained. The multiblock technique consists of dividing the physical domain into parts and, on each of the parts, applying the generation of meshes in generalized coordinates. Such a procedure allows to better capture the complexity of the geometry compared to the strategy employed on a single block (CIRILO, 2006; PARDO et al., 2012, SAITA et al., 2017).

\subsection{Monoblock Mesh Generation}
\label{sec3.2}

As a first application of the mesh generation algorithm, the set of input data presented in Table \ref{tab1} was used. The weighted values for $\Gamma_{j}$, $j =$ 1 and 3 are omitted, since they are known for complementarity, that is, $p^{\Gamma_{1} \Gamma_{3}}_{x} + p^{\Gamma_{2} \Gamma_{4}}_{x} = 1$ e $p^{\Gamma_{1} \Gamma_{3}}_{y} + p^{\Gamma_{2} \Gamma_{4}}_{y} = 1$.

\begin{table}[H]
	\begin{center}
		\caption[caption]{Input data of a Rectangular Geometry with Cutout}
		\begin{tabular}{c|c|c|c} \hline
			Number of partitions ($d^{f}$) & \multicolumn{2}{c|}{Number of partitions ($d^{t}$)} & Weighted average \\
			Input data & Refinement 1 & Refinement 2 & Weights ($p^{\Gamma_{2} \Gamma_{4}}$) \\ \hline
			$x \quad \quad y$ & $\xi \quad \quad \eta$ & $\xi \quad \quad \eta$ & $x \quad \quad y$ \\
			$10 \quad \quad 4$ & $15 \quad \quad 6$ & $30 \quad \quad 12$ & $0.9 \quad \quad 1.0$ \\ \hline
		\end{tabular}
		\label{tab1}
	\end{center}
\end{table}

Given the set of predefined points, Fig. 4a, the number of partitions and the weights at the borders, 
Table \ref{tab1}, then the border geometry is obtained by Parametric Spline Linear interpolation. In sequence, the algorithm calculates the new border points, using the given refinement. In this way, the algorithm starts to solve the transformation metrics, finally generating the mesh in generalized coordinates. Figs. 4b and 4c present the meshes in generalized coordinates, according to the refinements described in Table \ref{tab1}.

\begin{figure}[!ht]
\begin{center}
\caption{Rectangular Geometry with Cutout}
\subfigure[Predefined points]{\includegraphics[height=4.4cm,width=5.4cm]{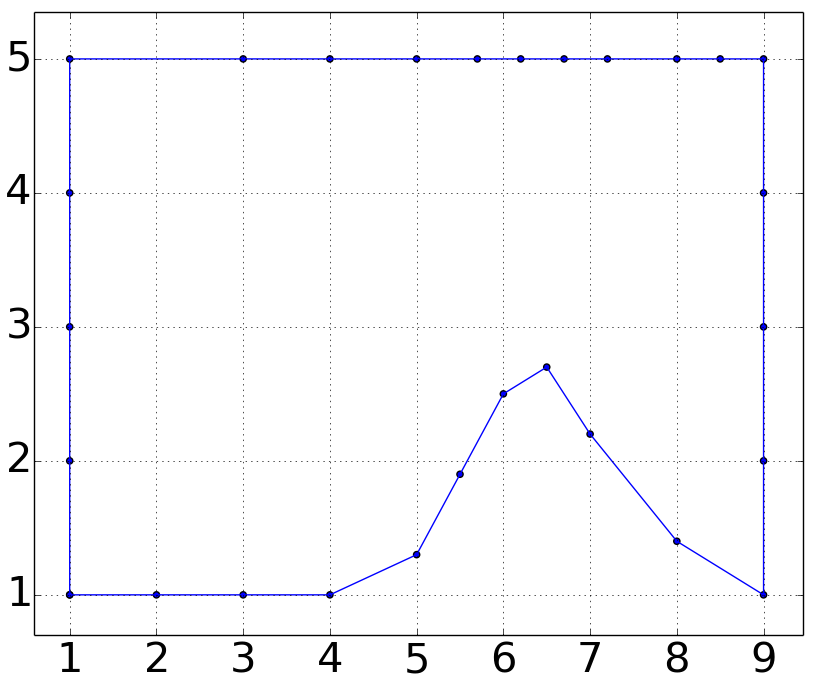}}
   \label{fig6a}\qquad \qquad
\subfigure[Refinement 1]{\includegraphics[height=4.4cm,width=5.4cm]{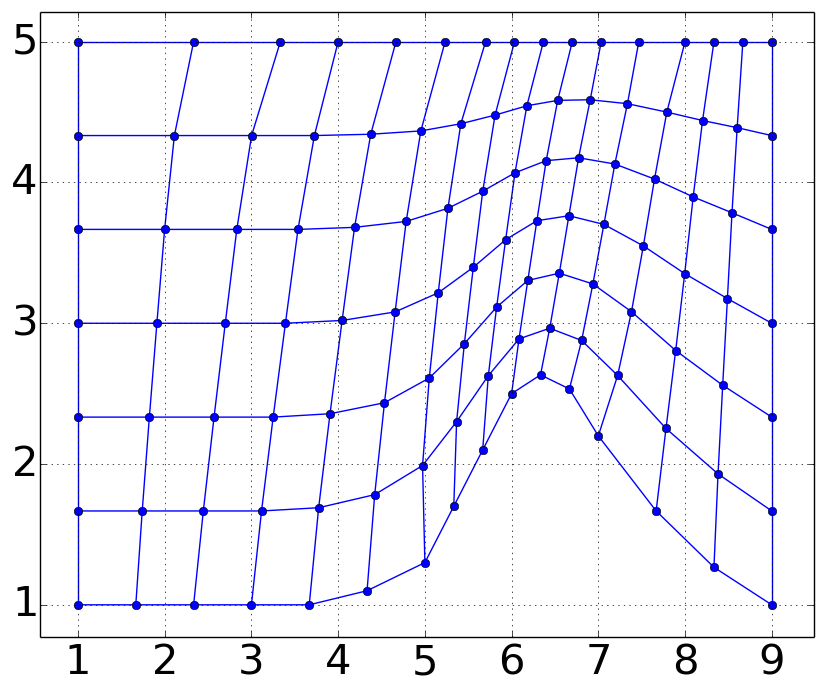}}
    \label{fig6b}\qquad \qquad
\subfigure[Refinement 2]{\includegraphics[height=4.4cm,width=5.4cm]{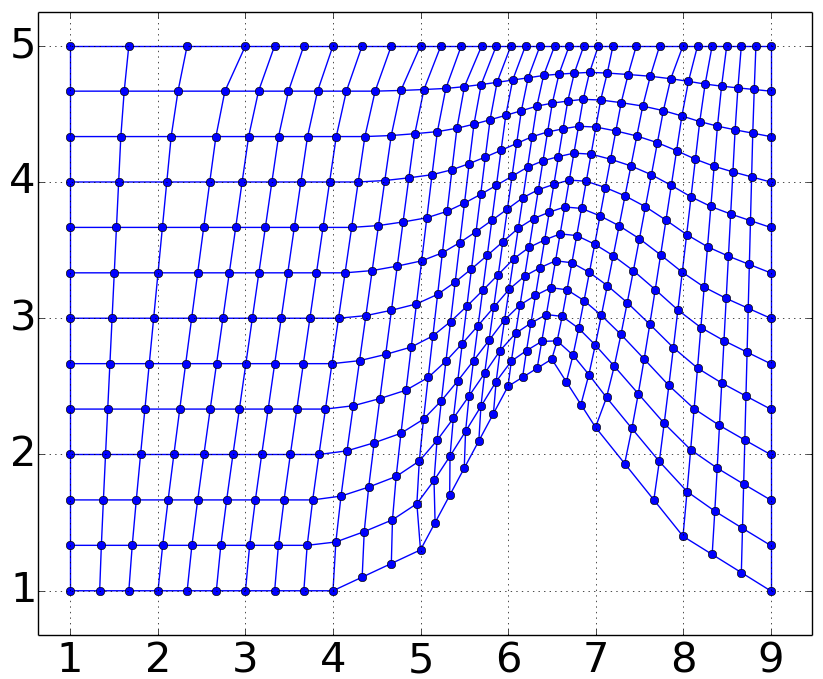}}
    \label{fig6c}
\end{center}
\center{{\bf{Source:}} The Authors}
\label{fig6}
\end{figure}

As a second application, a mesh in a doubly-connected domain was generated. The set of input data is shown in Table \ref{tab3}. 

\begin{table}[H]
	\centering
	\caption[caption]{Input data of a Doubly-Connected Geometry}
	\begin{tabular}{c|c|c|c} \hline
		Number of partitions ($d^{f}$) & \multicolumn{2}{c|}{Number of partitions ($d^{t}$)} & Weighted average \\
		Input data & Refinement 1 & Refinement 2 & Weights ($p^{\Gamma_{1} \Gamma_{3}}$) \\ \hline
		$x \quad \quad y$ & $\xi \quad \quad \eta$ & $\xi \quad \quad \eta$ & $x \quad \quad y$ \\
		$20 \quad \quad 4$ & $30 \quad \quad 6$ & $40 \quad \quad 8$ & $0.05 \quad \quad 0.0$ \\ \hline
	\end{tabular}
	\label{tab3}
\end{table}

In this application, the left/right border corresponds to the line segment that joins the two circles, while the upper/lower border corresponds to the two circles, as shown in Fig 5a. From the set of points given in Fig. 5a and the input data in Table \ref{tab3}, the algorithm constructs the meshes shown in Figs. 5b and 5c, according to the refinements described in Table \ref{tab3}.

\begin{figure}[!ht]
\begin{center}
\caption{Doubly-Connected Geometry}
\subfigure[Predefined points]{\includegraphics[height=4.4cm,width=5.4cm]{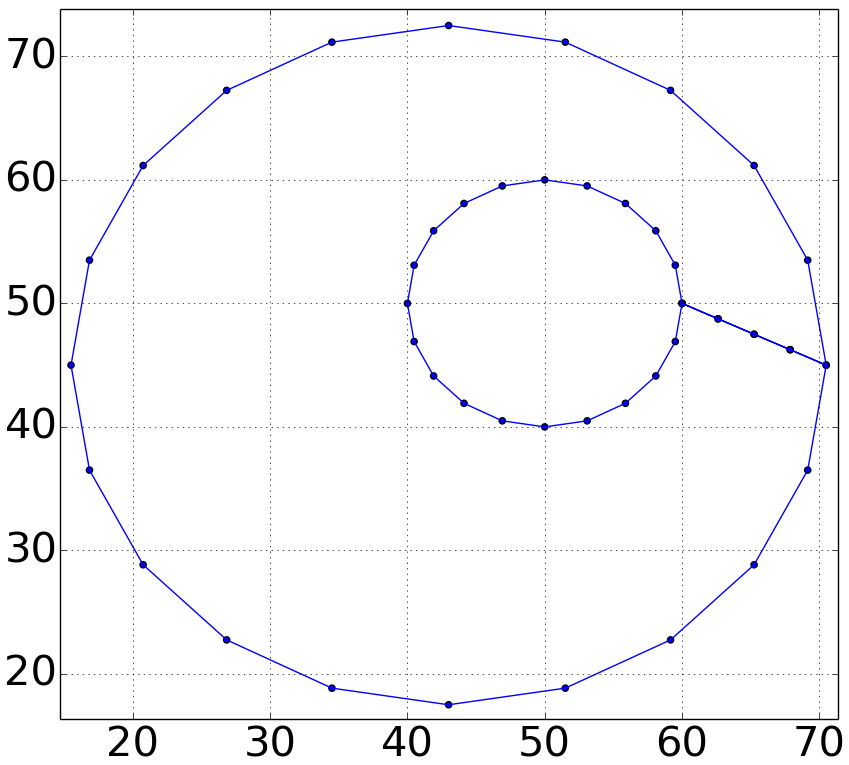}}
   \label{fig8a}\qquad \qquad
\subfigure[Refinement 1]{\includegraphics[height=4.4cm,width=5.4cm]{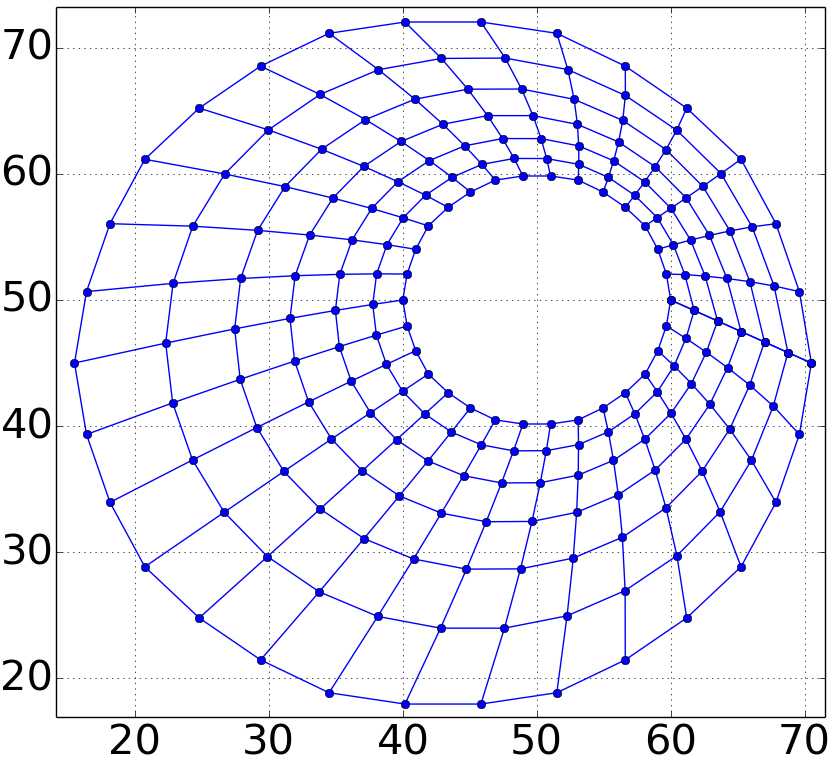}}
    \label{fig8b}\qquad \qquad
\subfigure[Refinement 2]{\includegraphics[height=4.4cm,width=5.4cm]{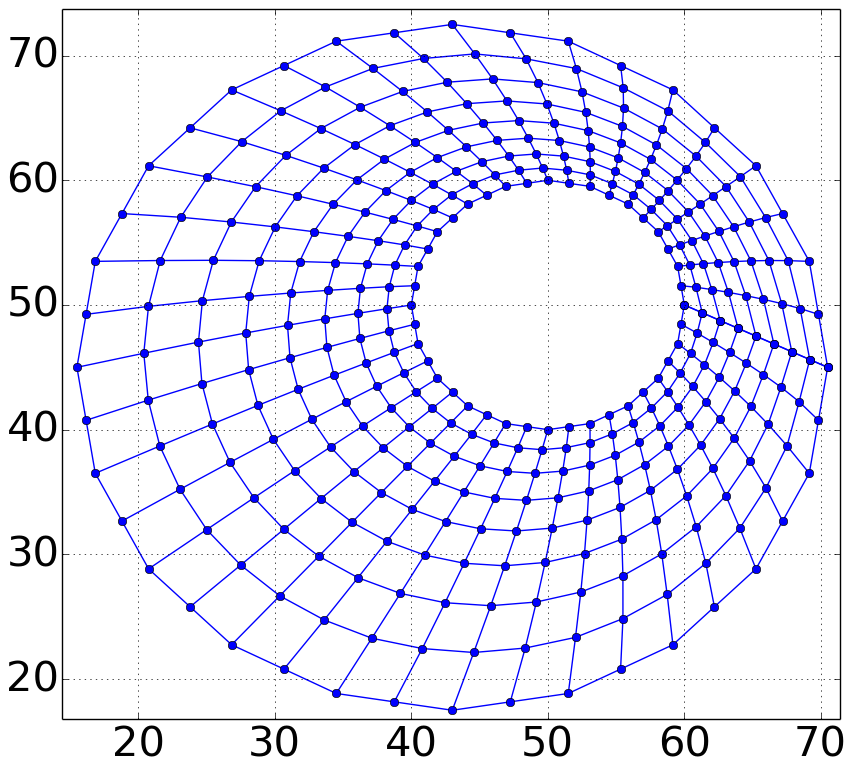}}
    \label{fig8c}
\end{center}
\center{{\bf{Source:}} The Authors}
\label{fig8}
\end{figure}

The latest generation of monoblock mesh is shaped like a bottle, as illustrated in Fig. 6. The set of input data is explained in Table \ref{tab2}.

\begin{table}[H]
	\centering
	\caption[caption]{Input data of a Bottle Geometry}
	\begin{tabular}{c|c|c|c} \hline
		Number of partitions ($d^{f}$) & \multicolumn{2}{c|}{Number of partitions ($d^{t}$)} & Weighted average \\
		Input data & Refinement 1 & Refinement 2 & Weights ($p^{\Gamma_{1} \Gamma_{3}}$) \\ \hline
		$x \quad \quad y$ & $\xi \quad \quad \eta$ & $\xi \quad \quad \eta$ & $x \quad \quad y$ \\
		$4 \quad \quad 10$ & $6 \quad \quad 15$ & $12 \quad \quad 30$ & $1.0 \quad \quad 0.5$ \\ \hline
	\end{tabular}
	\label{tab2}
\end{table}

Using the set of predefined points, Fig. 6a, and the input data in Table \ref{tab2}, the meshes in generalized coordinates are generated, as shown in Fig. 6b and Fig. 6c, according to the refinements described in Table \ref{tab2}.

\begin{figure}[!ht]
\begin{center}
\caption{Bottle Geometry}
\subfigure[Predefined points]{\includegraphics[height=4.4cm,width=5.4cm]{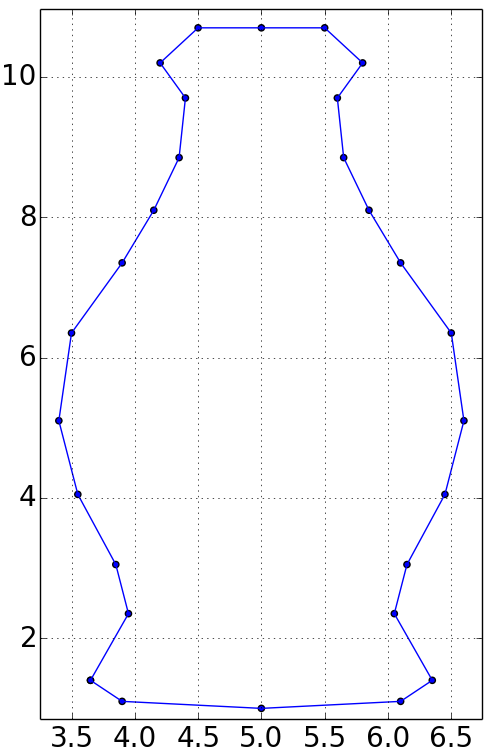}}
   \label{fig7a}\qquad \qquad
\subfigure[Refinement 1]{\includegraphics[height=4.4cm,width=5.4cm]{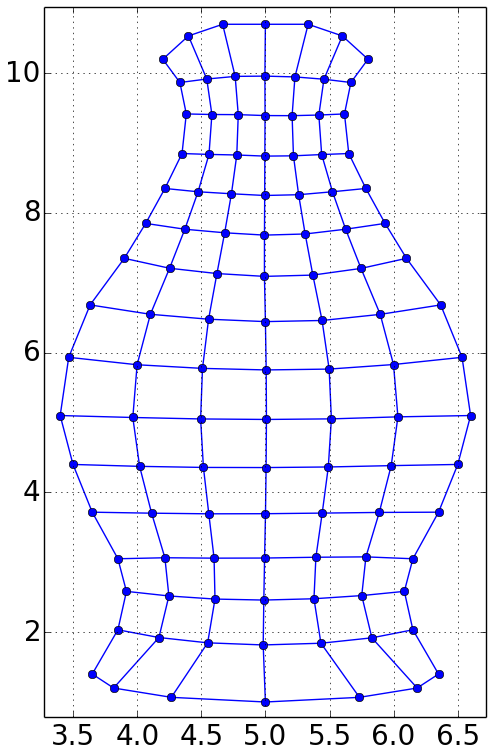}}
    \label{fig7b}\qquad \qquad
\subfigure[Refinement 2]{\includegraphics[height=4.4cm,width=5.4cm]{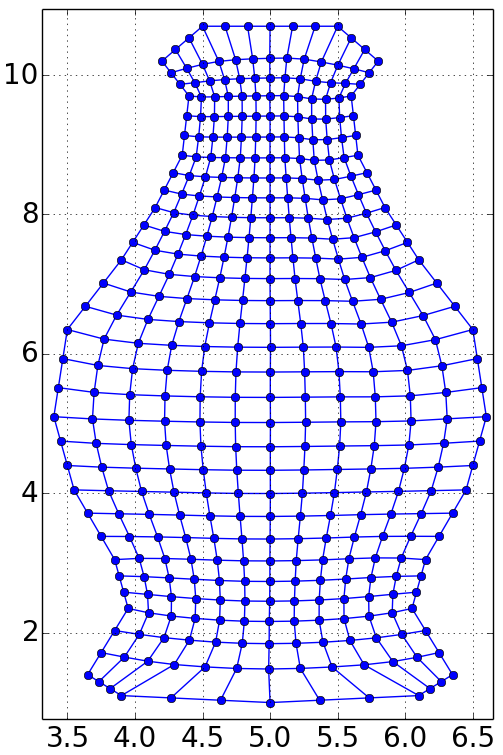}}
    \label{fig7c}
\end{center}
\center{{\bf{Source:}} The Authors}
\label{fig7}
\end{figure}

automated mesh generator
In the applications presented for monoblock meshes, the automated mesh generator precisely built the geometry in the physical domain.

\subsection {Multiblock Mesh Generation}

In order to show a more complex application, points from the contour of Lake Igapó I, located in Londrina, Paraná, Brazil, were collected. The points were obtained from satellite images offered by the Google Earth software (GOOGLE, 2020). The generated mesh has the same scale as the physical domain.

Unlike previous applications, in addition to generating the monoblock meshes (now called monobloc sub-meshes), in this application the sub-meshes must be connected. The multiblock mesh algorithm uses the monoblock mesh generator as a subroutine of the multiblock algorithm. 
\vspace {0.3cm}

In summary, the multiblock algorithm can be illustrated in this way:

\noindent
Step 1. Loop condition: If there is a monoblock input file being passed as an argument, perform step 2. Otherwise, perform step 4.

\noindent
Step 2. Monoblock mesh generation: Use the data present in the input file to generate the monoblock mesh.

\noindent
Step 3. Plot the monoblock mesh: Use the results of the monoblock mesh generator algorithm to plot the monoblock mesh, and return to step 1.

\noindent
Step 4. Plot the multiblock mesh: Show all monoblock meshes generated in a single multiblock mesh.
\vspace {0.3cm}

For the geometry of Lago Igapo, it is considered that it consists of three blocks, Lake Igapó I (main block) and its two tributaries (two other blocks). The set of predefined points are given in Fig. 7a. The input data are presented in Table \ref{tab4}. As for the weighted average, the percentage weights ($p^{\Gamma_{1} \Gamma_{3}}$ e $p^{\Gamma_{2} \Gamma_{4}}$), in $ x $ and $ y $, were all chosen as $ 0.5 $.

\begin{table}[H]
	\centering
	\caption[caption]{Input data of Igapó I Lake}
	\begin{tabular}{c|c|c|c} \hline
		Monoblock & Number of partitions ($d^{f}$) & \multicolumn{2}{c}{Number of partitions ($d^{t}$)} \\
		Name & Input data & Refinement 1 & Refinement 2 \\ \hline
		Principal & $x \quad \quad y$ & $\xi \quad \quad \eta$ & $\xi \quad \quad \eta$ \\		
		          & $6 \quad \quad 70$ & $6 \quad \quad 70$ & $12 \quad \quad 140$ \\ \hline
    Tributary 1 & $x \quad \quad y$ & $\xi \quad \quad \eta$ & $\xi \quad \quad \eta$ \\
                	& $10 \quad \quad 2$ & $10 \quad \quad 2$ & $20 \quad \quad 4$ \\ \hline
	  Tributary 2 & $x \quad \quad y$ & $\xi \quad \quad \eta$ & $\xi \quad \quad \eta$ \\
      	        	& $8 \quad \quad 5$ & $8 \quad \quad 5$ & $28 \quad \quad 10$ \\ \hline
		\end{tabular}
	\label{tab4}
\end{table}

After executing the multiblock mesh algorithm, the meshes were obtained in generalized coordinates as shown in Figs. 7b and 7c, according to the refinements of Table \ref{tab4}.

\begin{figure}[!ht]
\begin{center}
\caption{Igapó I Lake Geometry}
\subfigure[Predefined points]{\includegraphics[height=4.4cm,width=5.4cm]{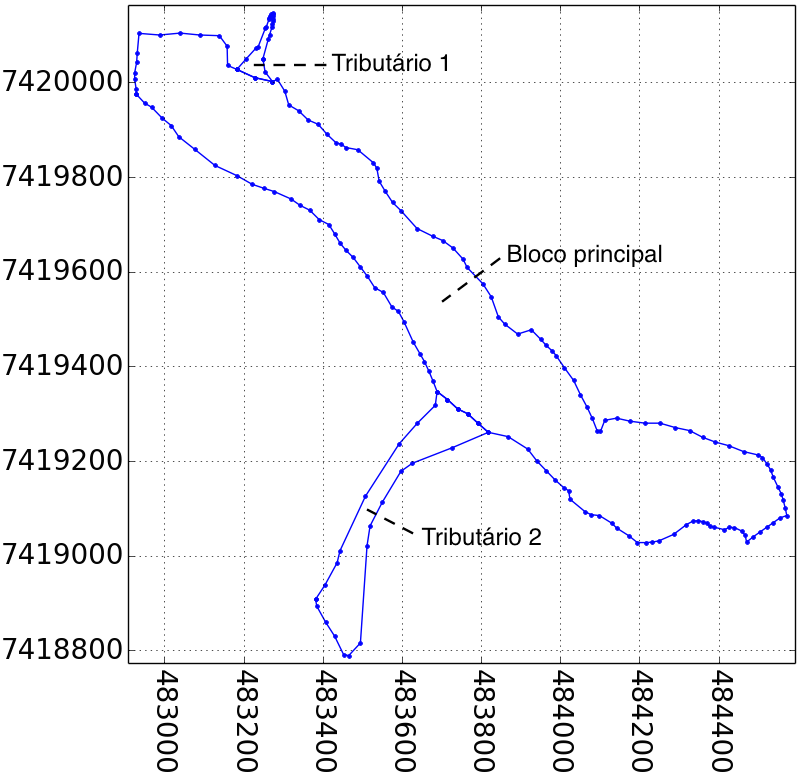}}
   \label{fig9a}\qquad \qquad
\subfigure[Refinement 1]{\includegraphics[height=4.4cm,width=5.4cm]{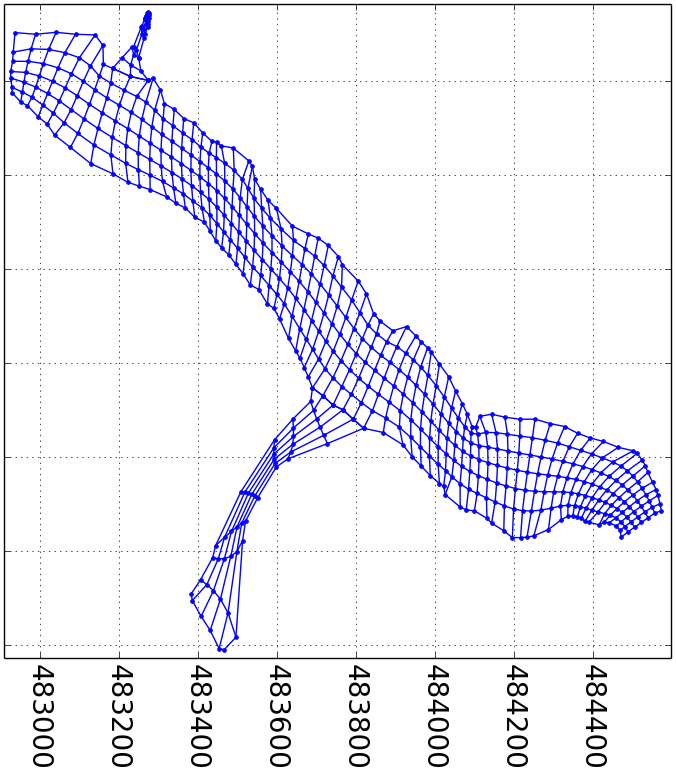}}
    \label{fig9b}\qquad \qquad
\subfigure[Refinement 2]{\includegraphics[height=4.4cm,width=5.4cm]{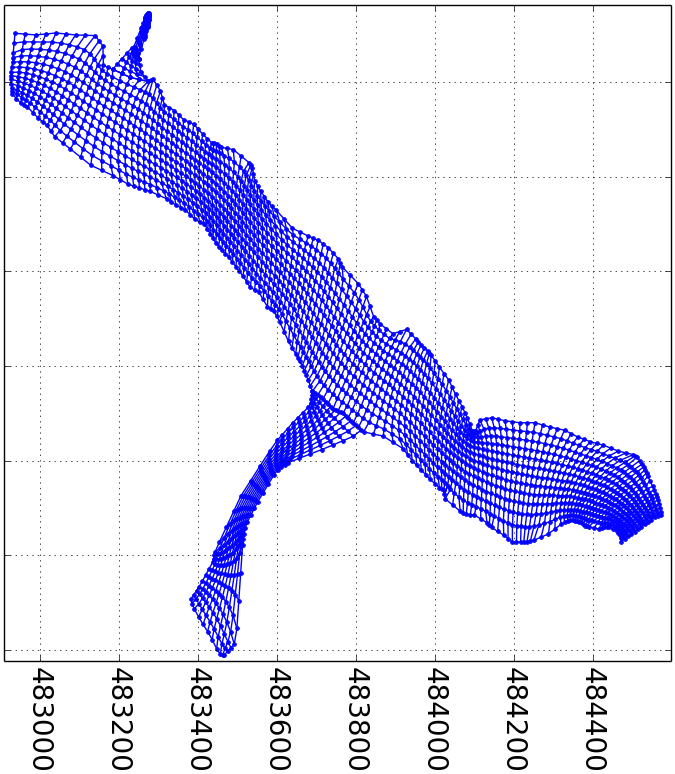}}
    \label{fig9c}
\end{center}
\center{{\bf{Source:}} The Authors}
\label{fig9}
\end{figure}

Again, the automated mesh generator accurately represented the physical domain of Igapó I lake, located in Londrina, Paraná, Brazil.

\section{Final Considerations}

In this work, using the generalized coordinate method, a mesh generator in Python was created and implemented. This automated mesh generator allowed the reproduction of complex geometries with several mesh refinements, as shown by the applications presented.

Specifically, as original contributions to this work, we highlight two aspects:

1- Parameterized Linear Spline. The use of the Parameterized Linear Spline method to obtain points at the border of the physical domain, given predefined points, allows the construction of the internal mesh in generalized coordinate, with the desired refinement, in a simple and fast way.

2- Python language. The Python programming language and the libraries used make it possible to optimize the algorithm, to facilitate the visualization of the meshes, and the portability of the application in different software architectures, such as Linux and Windows.

\vspace{0.9cm}

\noindent \textbf{\large{Referências}}

\vspace{0.5cm}

\noindent BELINELLI, E.O.; NATTI, P.L.; ROMEIRO, N.M.L.; CIRILO, E.R.; FANTIN, L.H.; OLIVEIRA, K.B.; CANTERI, M.G.; NATTI, E.R.T. {\it Geração de malha para descrever a dispersão da ferrugem da soja no Paraná}. In: Júlio Cesar Ribeiro. (Org.). Ciências Exatas e da Terra: Conhecimentos Estratégicos para o Desenvolvimento do País. 1ed.: Atena Editora, 2020, v. , p. 225-239.

\noindent BURDEN, R.L., FAIRES, J.D., BURDEN, A.M. \textit{Numerical Analysis}. Boston: Cengage Learning, 2015.

\noindent CAI, X., LANGTANGEN, H. P., MOE, H. On the performance of the Python programming language for serial and parallel scientific computations. \textit{Scientific Programming}, vol. 13, n. 1, pp. 31-56, 2005.

\noindent CIRILO, E.R.; DE BORTOLI, A.L. Cubic splines for trachea and bronchial tubes grid generation. {\it Semina: Exact and Technological Sciences}, v. 27, p.147-155, 2006.

\noindent CIRILO, E. R. ; BARBA, A. N. D. ; NATTI, P. L. ; ROMEIRO, N. M. L. . A numerical model based on the curvilinear coordinate system for the MAC method simplified. SEMINA. CIÊNCIAS EXATAS E TECNOLÓGICAS (ONLINE), v. 39, p. 87, 2018.

\noindent DE BORTOLI, A. L. \textit{Introdução à dinâmica de fluidos computacional}. Porto Alegre: Editora da UFRG, 2000.

\noindent FORTUNA, A.O. \textit{Computational techniques for fluid dynamics: Basic concepts and applications}. S\~ao Paulo: EDUSP, 2012.

\noindent GOOGLE. \textit{Google Earth}. Disponível em: <\underline{\url{https://www.google.com/earth/}}>. Acessed in oct. 2020.

\noindent KOOMULLIL, R.; SONI B.; SINGH R. A comprehensive generalized mesh system for CFD applications. \textit{Mathematics and Computers in Simulation}, vol. 78, pp. 605-617, 2008.

\noindent LAIPING, Z.; ZHONG, Z.; XINGHUA, C., XITHAMES, H. A 3D hybrid grid generation technique and a multigrid/parallel algorithm based on anisotropic agglomeration approach. \textit{Chinese Journal of Aeronautics}, vol. 26, n.1, pp. 47-62, 2013.

\noindent MALISKA, C. R. \textit{Transferência de calor e mecânica dos fluidos computacional}. São Paulo: LTC, 2004.

\noindent PARDO, S.R.; NATTI, P.L.; ROMEIRO, N.M.L.; CIRILO, E.R. A transport modeling of the carbon-nitrogen cycle at Igap\'o I Lake-Londrina, Paran\'a State, Brazil. {\it Acta Scientiarum. Technology}, v. 34, p. 217-226, 2012.

\noindent ROMEIRO, N.L.M.; CASTRO, R.G.S.; CIRILO, E.R.; NATTI, P.L. Local calibration of coliforms parameters of water quality problem at Igap\'o I Lake - Londrina, Paran\'a, Brazil. {\it Ecological Modelling}, v. 222, p. 1888-1896, 2011.

\noindent ROMEIRO, N.L.M.; MANGILI, F.B.; COSTANZI, R.N.; CIRILO, E.R.; NATTI, P.L. Numerical simulation of BOD5 dynamics in Igap\'o I lake, Londrina, Paran\'a, Brazil: Experimental measurement and mathematical modeling. {\it Semina: Exact and Technological Sciences}, v. 38, p. 50-58, 2017.

\noindent SAITA, T.M.; NATTI, P.L.; CIRILO, E.R.; ROMEIRO, N.L.M.; CANDEZANO, M.A.C.; ACUNA, R.A.B.; MORENO, L.C.G. Simula\c{c}\~ao num\'erica da din\^amica de coliformes fecais no lago Luruaco, Col\^ombia. {\it Trends in Applied and Computational Mathematics}, v. 18, p. 435-447, 2017.	

\noindent SCIPY-NUMPY. In <\underline{\url{http://www.numpy.org/}}>. Acessed in oct. 2020.

\noindent SCIPY-MATPLOTLIB. In <\underline{\url{http://matplotlib.org/}}>. Acessed in oct. 2020.

\noindent THOMPSON, J. F.; THAMES, F. C.; MASTIN, C. W. TOMCAT - A Code for Numerical Generation of Boundary Fitted Curvilinear Coordinate Systems on Fields Containing Any Number of Arbitrary Two-Dimensional Bodies. \textit{Journal of Computational Physics}, vol. 24, pp. 274-302, 1977.

\noindent THOMPSON, J. F.; WARSI, Z. U. A.; MASTIN, C. W. \textit{Numerical grid generation: foundations and applications}. New York: Elsevier Science Publishing, 1985.

\noindent THOMPSON, J.F.; SONI, B.K.; WEATHERILL, N.P. \textit{Handbook of Grid Generation}. Florida: CRC Press, 1998.

\end{document}